\documentclass[aps,prl,twocolumn]{revtex4}
\pdfoutput=1
\usepackage{graphicx}
\usepackage{amssymb}
\usepackage{amsmath}
\usepackage{bm}
\usepackage{color}

\makeatletter

\begin{document}

\title{Stability of persistent currents in open-dissipative quantum fluids}

\author{Guangyao Li$^1$, Michael D. Fraser$^2$,  Alexander Yakimenko$^3$, and Elena A. Ostrovskaya$^1$}

\affiliation{$^1$Nonlinear Physics Centre, Research School of Physics and Engineering, The Australian National University, Canberra ACT 0200, Australia,  $^2$Quantum Optics Research Group, RIKEN Center for Emergent Matter Science, 2-1 Hirosawa, Wako-shi, Saitama 351-0198, Japan, \\ $^3$Department of Physics, Taras Shevchenko National University, Kyiv 01601, Ukraine }


\begin{abstract}
The phenomenon of stable persistent currents is central  to the studies of superfluidity in a range of physical systems. While all of the previous theoretical studies of superfluid flows in annular geometries concentrated on conservative systems, here we extend the stability analysis of persistent currents to open-dissipative exciton-polariton superfluids. By considering an exciton-polariton condensate in an optically-induced annular trap, we determine stability conditions for an initially imposed flow with a non-zero orbital angular momentum. We show, theoretically and numerically, that the system can sustain metastable persistent currents in a large parameter region, and describe scenarios of the supercurrent decay due to the dynamical instability.
\end{abstract}
\maketitle

{\em Introduction.} -- Superfluidity, which is an ability of a fluid to flow without friction,
has been studied in various systems including the superfluid helium \cite{Avenel85,Varoquaux86,Davis92},
superconductivity \cite{Annett05}, Bose-Einstein condensate  (BEC) of dilute atomic gases \cite{Pitaevskii03},
and, more recently, exciton-polariton BECs in semiconductor microcavities \cite{Deng10,Carusotto13}.

One of the most important predictions of quantum hydrodynamics is the formation of persistent currents of a superfulid confined in an annular trap with an initially imposed rotation. Apart from the fundamental interest in this problem, ultra-sensitive interferometric  devices based on persistent currents have been suggested \cite{SQUID_pBEC,SQUID_aBEC1,SQUID_aBEC2}.  The ability to use the Laguerre-Gaussian (LG) mode of an optical laser to trap atomic BECs and directly transfer orbital angular momentum from photons to atoms \cite{Ryu07,SQUID_aBEC1,Moulder12} fuelled  intensive studies of persistent currents in atomic condensates. Stability analysis \cite{Javanainen98, Salasnich99,Javanainen01,Bargi10,Modugno04}, confirmed that persistent currents are metastable states with lifetimes limited only by longevity of the BEC \cite{Moulder12}. Although previous theoretical results agree with experiments on ultracold atomic gases, their scope was limited to conservative systems at thermal equilibrium.

\begin{figure}
\includegraphics[width=8cm]{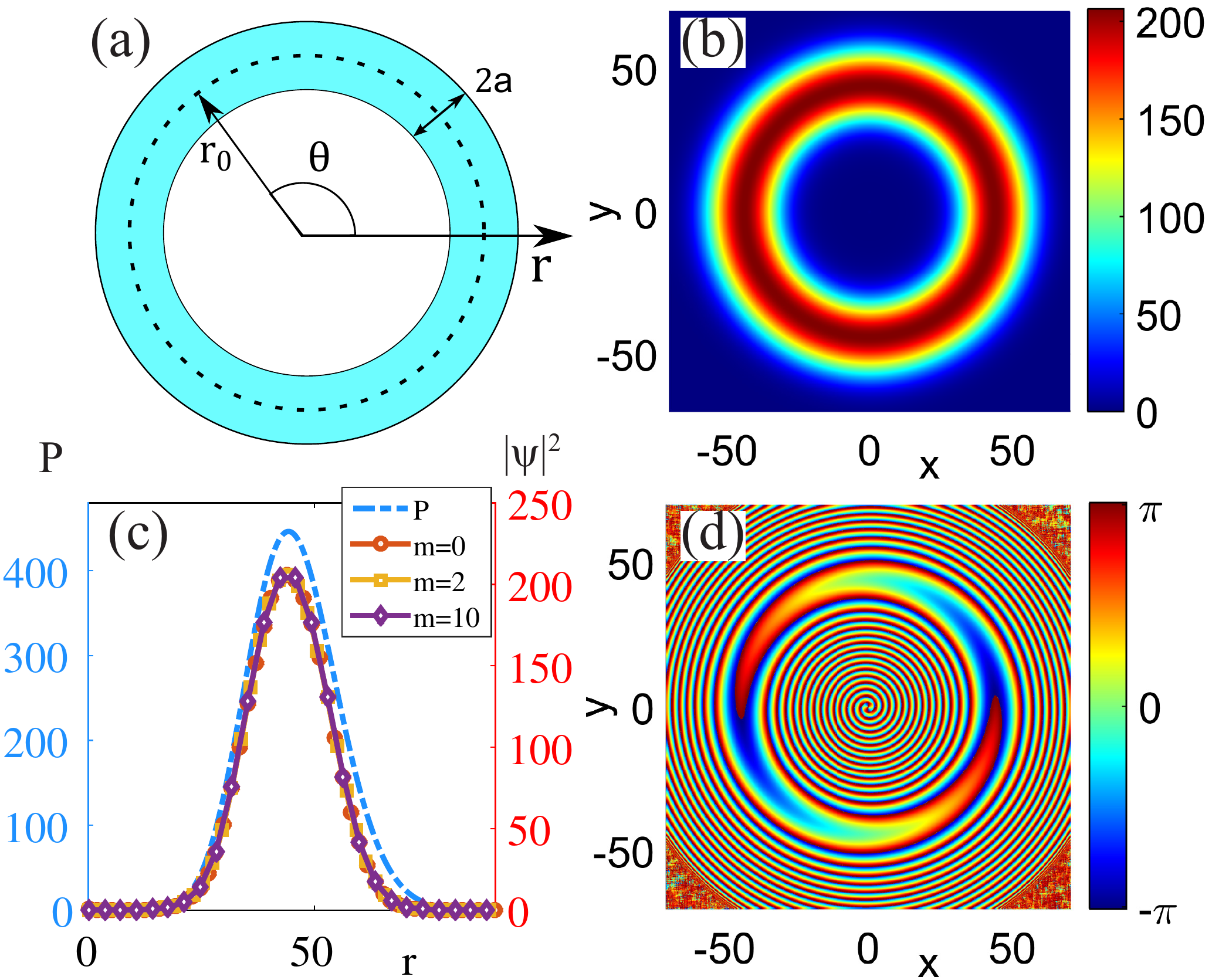} 
\caption{(Color online). (a) Schematics of the radial structure of a LG pump beam; Steady state (b) density and (d) phase of the condensate for $m=2$; (c) Radial profiles of the steady states with $m=0,2,10$ supported by the LG$_{50}$ mode.}
\label{Steady_States}
\end{figure}

Applicability of the existing theories to the novel quantum fluids formed by exciton-polariton condensates is questionable. Polaritons are quasiparticles arising from strong coupling between photons confined in a microcavity and excitions in a quantum well \cite{Deng10,Carusotto13}. The polariton condensates can be generated either by coherent (resonant) or incoherent (off-resonant) optical pumping schemes. While the former leads to a  condensate which is driven directly by the pumping laser \cite{Savvidis00}, the later relies on non-radiative relaxation processes in the microcavity and stimulated scattering into the lowest energy state, which leads to a condensate with spontaneously established coherence \cite{Huang00}. Regardless of the excitation scheme, and in contrast to ultracold atomic gasses, an exciton-polariton BEC is an intrinsically non-equilibrium system because of the pumping and radiative decay of polaritons. With the rapid growth of interest in  persistent flows of non-equilibrium, open-dissipative polariton condensates \cite{Toroidal_pBEC_11,Snoke14,Toroidal_pBEC}, the urgent open question is how the intrinsic gain and loss would affect their stability.

In this work, we address this problem by constructing a comprehensive theory of polariton condensates with non-zero orbital angular momenta supported by an {\em optically induced annular confinement}. We focus on the incoherent, off-resonant pumping scheme which offers the possibility to engineer a trapping potential landscape by shaping the optical pump beam \cite{gain_guiding,sculpting_osc, Cristofolini13,Askitopoulos13} and ensures that the condensate's phase evolution does not depend on the phase of the pump. We predict that persistent currents of polaritons with sufficiently high angular momentum are always prone to {\em oscillatory} dynamical instabilities. However,  in sizeable regions of the parameter space, the quantized circulation can persist almost indefinitely.

{\em The steady currents.} -- The non-resonantly pumped polariton condensate can be described by the mean-field dissipative Gross-Pitaevskii equation for the macroscopic wavefunction, $\psi$, coupled to the rate equation for the density of the excitonic reservoir, $n_R$ \cite{Wouters07}:
\begin{equation}\label{GPE}
\begin{aligned}
i \hbar\frac{\partial \psi}{\partial t}=&\left\{-\frac{\hbar^2}{2m}\nabla^2+g_c |\psi|^2+g_R n_R +\frac{i \hbar}{2}\left[ R\, n_R-\gamma_c \right] \right \} \psi, \\
\frac{\partial n_R}{\partial t}=&P-\left(\gamma_R+R |\psi|^2\right) n_R, 
\end{aligned}
\end{equation}
where $P(\textbf{r},t)$ is the pumping rate, $g_c$ and $g_R$ characterise polariton-polariton and polariton-exciton interactions, respectively. The relaxation rates $\gamma_c$ and $\gamma_R$ quantify the finite lifetime of condensed polaritons and the reservoir, respectively. The stimulated scattering rate,  $R$, controls growth of the condensate density. In what follows, we will consider the dimensionless form of  Eq.~(\ref{GPE}) obtained by introducing the characteristic time $T=\gamma_c^{-1}$, length $L=\sqrt{\hbar/(m\gamma_{c})}$, and energy $E=\hbar \gamma_c$ scales \cite{units}.

Optical trapping techniques \cite{gain_guiding,sculpting_osc, Cristofolini13,Askitopoulos13, Toroidal_pBEC_11,Toroidal_pBEC,Snoke14} rely on effective trapping potentials for polaritons created due to polariton flows and interaction with the reservoir. In the spirit of this approach,  the annular condensate can be supported by the LG pump beam. The angular momentum carried by the LG beam will not be transferred to the condensate because polaritons lose coherence in the reservoir during the scattering between "hot" reservoir polaritons and phonons, and the intensity of the LG beam will define the spatial distribution of the condensed state.  Within the homogeneous approximation, the threshold of pumping power to build up a non-zero condensate density is $P_{th}=\gamma_R \gamma_c / R$ \cite{Wouters07}, and this value can be used to normalise intensity distribution of the LG mode as $\bar{P}(r)=P(r)/P_{th}$. Equations~(\ref{GPE}) with the radially symmetric pump admit steady states of the form: $\psi=\Psi(r,\theta)\exp(im\theta)\exp(-i\mu t)$, where $\mu$ is the energy (chemical potential) of the steady state, $(r,\theta)$ are the polar coordinates, and $m$ is the phase winding number \cite{anton} (topological charge of a vortex) with the ground state corresponding to $m=0$. Such steady states can be found by solving (\ref{GPE}) numerically, with the initially imposed vorticity  $\psi(0)=Ar^{|m|}\exp(im\theta)$, and some examples for $m=0$ and $m\neq0$ are presented in Fig. \ref{Steady_States}. Remarkably, the radial profiles of the condensate show extremely weak dependence on $m$  [Fig. \ref{Steady_States} (c)]. In experiment, the initial vorticity can be imprinted, e.g., by a pulsed resonant transfer of the orbital angular momentum onto an established $m=0$ state \cite{Sanvitto10}.
 
Within the pump area quantised superfiuld flows are supported purely by the balance of gain and loss, and therefore resemble {\em dissipative vortex solitons} in a focusing optical media \cite{Borovkova12}. At the same time, the steady state maintains inward and outward polariton flows outside the pump area, which creates an effective trapping potential in the radial direction \cite{Ostrovskaya12,Ge14}. If the inward polariton flow does not decay fast enough, then polaritons might form a central density peak \cite{Cristofolini13,Askitopoulos13}. This phenomenon will not occur in non-zero angular momentum states \cite{Toroidal_pBEC}, since the phase singularity is associated with a vanishing density at the vortex core \cite{Pitaevskii03}. 

In numerical simulations, the steady states are characterised by the conserved real part of the full energy functional, $E_0$, and angular momentum, $L_z$:
\begin{equation}\label{Energy}
E_0=\, \int \left[ \frac{1}{2} |\nabla \psi|^2+g_R\, n_R \, |\psi|^2+\frac{g_c}{2}|\psi|^4 \right] rdr d\theta,
\end{equation} 
\begin{equation}\label{Momentum}
L_z=\frac{i}{2}\int \left(\psi\frac{\partial \psi^*}{\partial \theta}-\psi^*\frac{\partial \psi}{\partial \theta}\right) \, rdrd\theta.
\end{equation}
For a steady state with azimuthally homogeneous density, the normalised angular momentum is equal to the topological charge of the vortex: $L_z/\int |\psi|^2 r dr d\theta=m$.

{\em 1D approximation. --}  When the radius of the LG beam is much larger than the width of the annulus, \emph{i.e.}, $r_0 \gg a$, one can separate the radial and azimuthal dependence of the condensate wavefunction  \cite{Rokhsar97,Javanainen98,Bargi10,Stringari06,Baharian13} and derive a one-dimensional model, which was shown to agree with its higher-dimensional counterparts in the conservative case \cite{Salasnich99}. To this end, we set $\psi(r,\theta)=\Phi_0(r)\psi(\theta,t)$, where $r \in [r_0-a, r_0+a]$ and $\Phi_0(r)$ is assumed to take a constant value over the width of the ring, $2a$.  Substituting this ansatz into our model, and integrating out the radial dependence, we arrive at the reduced 1D model:
\begin{equation}\label{GPE_1D}
\begin{aligned}
i \frac{\partial \psi (\theta,t)}{\partial t}=& \left\{-\frac{1}{2 r^2_0}\frac{\partial^2}{\partial \theta^2}+g_c n^0_c|\psi(\theta,t)|^2+
g_R n_R(\theta,t) \right. \\
                                                           &\;\;\left. +\frac{i}{2}\left[ R n_R(\theta,t)-\gamma_c \right] \right \} \psi(\theta,t), \\
   \frac{\partial n_R(\theta,t)}{\partial t}=&\;P(\theta)-\left(\gamma_R+R n^0_c|\psi(\theta,t)|^2\right) n_R(\theta,t), \\ 
\end{aligned}
\end{equation}
where, assuming our normalisation, $\gamma_c=1$. 

For a steady state, which is homogeneous in the radial direction, gain balances loss: $R n^0_R=\gamma_c$, were $n^0_R$ is the steady state reservoir density. The chemical potential of the stationary condensate with the azimuthal wave function $\psi(\theta,t)=\psi^0_\theta=\exp(im\theta)\exp(-i\mu t)$ is given by $\mu=m^2/(2 r^2_0)+ g_c n^0_c +g_R n_R^0$, where $n^0_c=\Phi^2_0=\gamma_R (\bar{P}-1)/R$ is the condensate density. 

{\em Excitation spectra.} -- Stability of the steady states with non-zero angular momentum can be analysed following the standard Bogoliubov-de Gennes (BdG) approach \cite{Pitaevskii03}, by calculating the spectrum of the elementary excitations of the condensate and the reservoir in our reduced one-dimensional model: $\psi(\theta,t)=\psi^0_\theta(1+\delta \psi)$, and $n_R(t)=n^0_R+\delta n_R$. The excitations of the steady state and its reservoir are introduced in the form \cite{Wouters07,Wouters10}:
\begin{equation}\label{delta_psi}
	\begin{aligned}
	\delta\psi=&\;u_0\, e^{ik\theta-i\omega t}+v_0^*\, e^{-ik\theta+i\omega^*t}\\
	\delta n  = &\;w_0\, e^{i k\theta-i\omega t}+w_0^*\, e^{-i k \theta+i \omega^* t}.
	\end{aligned}
\end{equation}
Inserting $\psi(\theta)$ and $n_R$ into Eq.~(\ref{GPE_1D}), and keeping only linear terms of $\mathcal{U}=(u_0,v_0,w_0)^T$, we obtain the BdG equations: $\mathcal{L}_m (k) \,\mathcal{U}=\omega \,\mathcal{U}$, where
\begin{equation}\label{L}
\mathcal{L}_m (k)=\begin{bmatrix}
(k_++g_c n^0_c) &            g_c n^0_c             &(g_R+\frac{i}{2}R) \\ 
-g_c n^0_c&                    (k_--g_c n^0_c)        &-(g_R-\frac{i}{2}R) \\
 -i\gamma_c n^0_c&    -i \gamma_c n^0_c     &-i (\gamma_R +R n^0_c)
\end{bmatrix}. \nonumber
\end{equation}
Here $k_\pm=\pm(1/2)(\bar{k}^2 \pm 2\bar{k}\bar{m})$ and $\{\bar{k}, \bar{m}\}=\{k/r_0,m/r_0\}$.

The spectrum of elementary excitations for $m=0$ is well  known  \cite{Carusotto13,Wouters07,Littlewood06,Malpuech14,smirnov}. For $m\neq 0$, the dispersion relation given by the BdG equations is:
\begin{multline}
\omega^3-\omega^2(2\bar{k}\bar{m}-i\bar{\gamma}_R)-\\ \omega(\omega^2_B-\bar{k}^2\bar{m}^2+R\bar{\gamma}_c-2i\bar{\gamma}_R\bar{k}\bar{m})-f(\bar{k},\bar{m})=0, \nonumber
\end{multline}
where $f(\bar{k},\bar{m})=\bar{\gamma}_c(ig_R\bar{k}^2+R\bar{k}\bar{m})+i\bar{\gamma}_R(\omega^2_B-\bar{k}^2\bar{m}^2)$,  $\omega^2_B=\bar{k}^4/4+g_c\Phi^2_0\bar{k}^2$ is the standard Bogoliubov dispersion, and we introduced the shorthand notations: $\bar{\gamma}_c=\gamma_c\Phi^2_0=P_{th}(\bar{P}-1)$ and $\bar{\gamma}_R=\gamma_R+R\Phi^2_0=\gamma_R\bar{P}$.

\begin{figure}[here]
\includegraphics[width=\columnwidth]{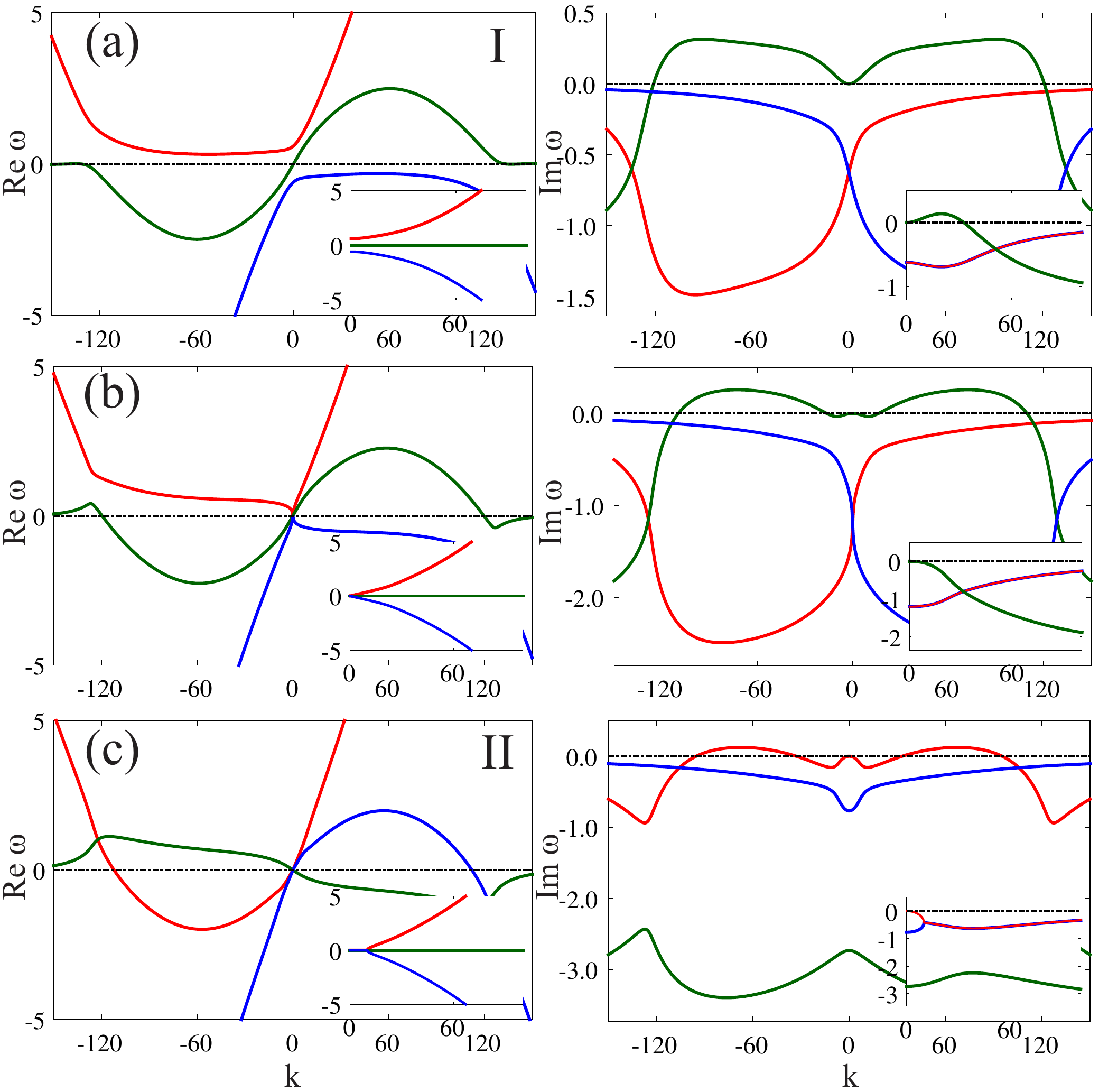}
\caption{(Color online). Dispersion curves ${\rm Re}[\omega(k)]=\Omega(k)$ and ${\rm Im}[\omega(k)]=\Gamma(k)$ for fixed $P_0=2.5$, $m=60$, and (a) $P_0<\gamma_c/\gamma_R$, (b)  $P_0=\gamma_c/\gamma_R$, (c) $P_0>\gamma_c/\gamma_R$. Insets show corresponding dispersion for $m=0$.}
\label{Full_dispersion}
\end{figure}
\begin{figure}[here]
\includegraphics[width=\columnwidth]{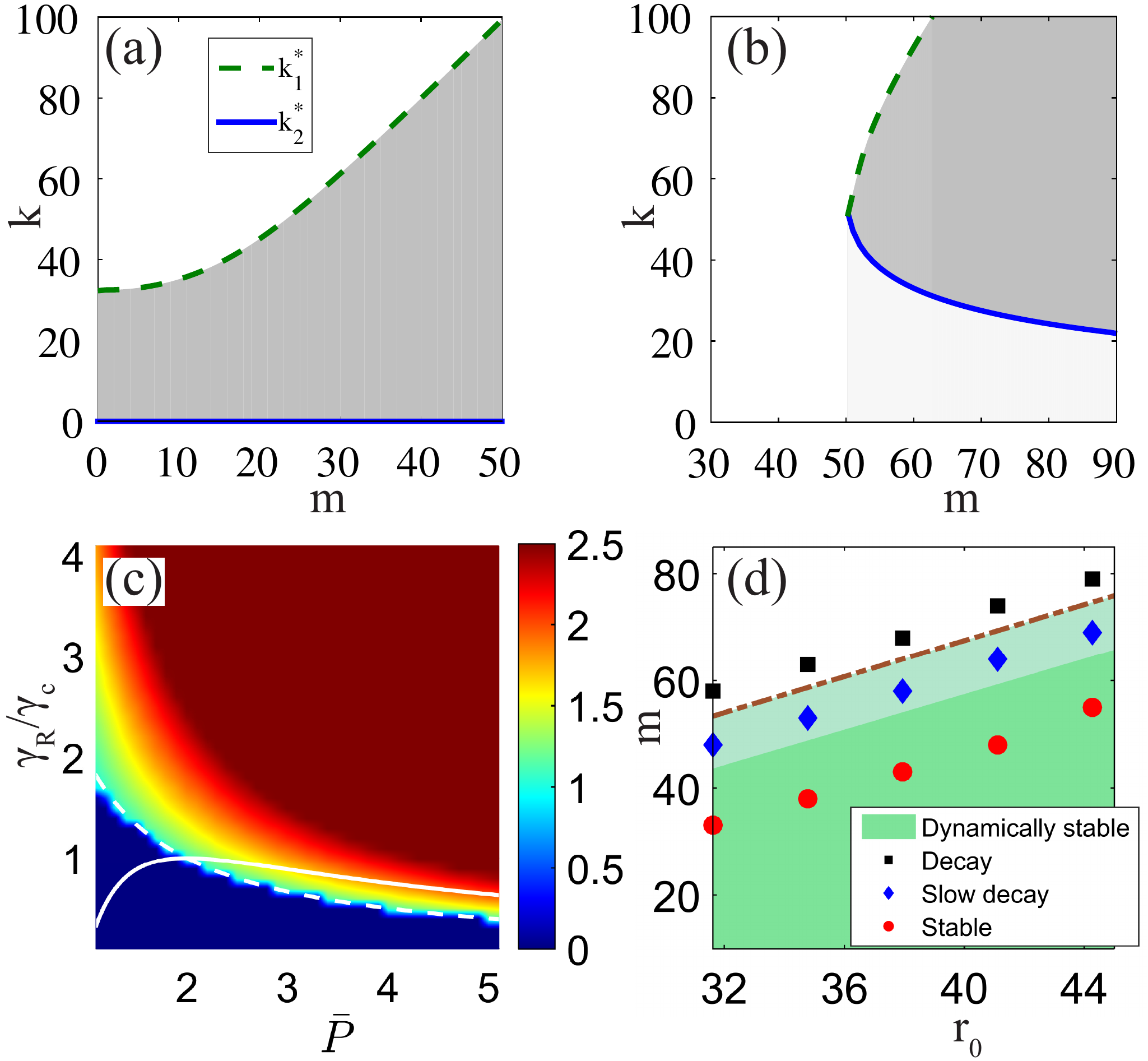}
\caption{(Color online). Critical values of $k^*_1$ and $k^*_2$ defining the modulational instability domains (shaded) in the regimes (a) $I$ and (b) $II$ corresponding to Fig. \ref{Full_dispersion} (a) and (c), respectively. (c) Phase diagram of $m_{ds}/r_0$. Dashed line, $\bar{P}=(g_R/g_c)(\gamma_c/\gamma_R)$, separates the MI regimes $I$ and $II$. Solid line is given by $P_0=\gamma_c/\gamma_R$ (see text); (d) Stability domains in the $m$ vs $r_0$ plane. Dots correspond to the 2D numerical simulations, the boundary of the dynamically stable (shaded) region, $m_{ds}$, is given by the 1D theory.}
\label{k_vs_m}
\end{figure}

At $k=0$, the real part of the excitation frequency $\omega(0)=\Omega(0)+i\Gamma(0)$ is found as $\Omega^2(0)=R\bar{\gamma_c}-\bar{\gamma}^2_R/4$. Consequently, it turns to zero for a critical pumping power $P_0\equiv\bar{P}^2/[4(\bar{P}-1)]=\gamma_c/\gamma_R$, and the spectrum near $k=0$ resembles the Bogoliubov dispersion. For $P_0<\gamma_c/\gamma_R$,  the Goldstone mode ($\omega=0$) at $k=0$ is separated from the non-zero mode by a gap of the size $\Omega^2(0)>0$ \cite{Byrnes12}. For $P_0>\gamma_c/\gamma_R$,  $\Omega^2(0)<0$ and the excitations exhibit a diffusive behaviour near $k=0$ \cite{{Littlewood06,Wouters07,WoutersPRA07}}. The gapped and diffusive character of the excitation spectra can be linked, respectively, to the underdamped and overdamped oscillations of the reservoir discussed in \cite{Byrnes12}. Figure~\ref{Full_dispersion} shows typical dispersion curves for the $m\neq0$ in the gapped (a) and diffusive (c) regimes and the marginal case $P_0=\gamma_c/\gamma_R$ [Fig. \ref{Full_dispersion}(b)]. 

{\em Dynamical stability.} -- When the imaginary part of the excitation frequency becomes positive, $\Gamma(k)>0$, for an interval $k^*_1<k<k^*_2$, the corresponding steady state experiences {\em modulational instability} (MI). As seen from Fig. \ref{Full_dispersion}, for $m\neq 0$ the corresponding real part of the excitation frequency is always non-zero, $\Omega(k)\neq 0$, which indicates the {\em oscillatory} nature of the instability. The polariton current exhibits MI only above certain critical angular momentum $m>m_{ds}$, which is defined by $\Gamma(k)$ crossing into the positive half-plane, at which point $k^*_1=k^*_2\neq0$. Two regimes of instability can be identified:

{\em Regime I} corresponds to $m_{ds}=0$ and is defined by the condition ${\bar P} <(g_R/g_c)(\gamma_c/\gamma_R)$ \cite{smirnov}. In this regime, the ground state $m=0$ is modulationally unstable, and $k^*_1=0$, as shown in the inset on the right panel of Fig. \ref{Full_dispersion}(a) and in Fig. \ref{k_vs_m}(a). The real part of the corresponding excitation frequency is zero, $\Omega(k)=0$ $(0<k<k^*_2)$, so that perturbations of the $m=0$ state grow exponentially and lead to fragmentation of the azimuthally homogeneous steady state. In this parameter range, due to the saturable nature of gain in this system, the effective nonlinearity becomes {\em attractive} for low condensate densities \cite{smirnov}. As seen from Fig. \ref{k_vs_m}(c) (below dashed line), this regime mostly overlaps with the gapped domain of the excitation spectra (below the solid line). Physically, this behaviour appears to be most relevant near the condensation threshold, $\bar P\sim 1$ due to the long lifetimes of the reservoir compared to condensate polaritons $\gamma_R/\gamma_c<1$.

{\em Regime II} corresponds to $m_{ds}>0$ and ${\bar P} >(g_R/g_c)(\gamma_c/\gamma_R)$ [Fig. \ref{k_vs_m}(c), above dashed line]. In this regime the ground state $m=0$ is dynamically stable, and the 1D theory predicts dynamical stability of the flow against azimuthal density modulations up to reasonably high values of $m_{ds}$ [Fig. \ref{k_vs_m}(b,c)]. 

\begin{figure}[h!]
\includegraphics[width=\columnwidth]{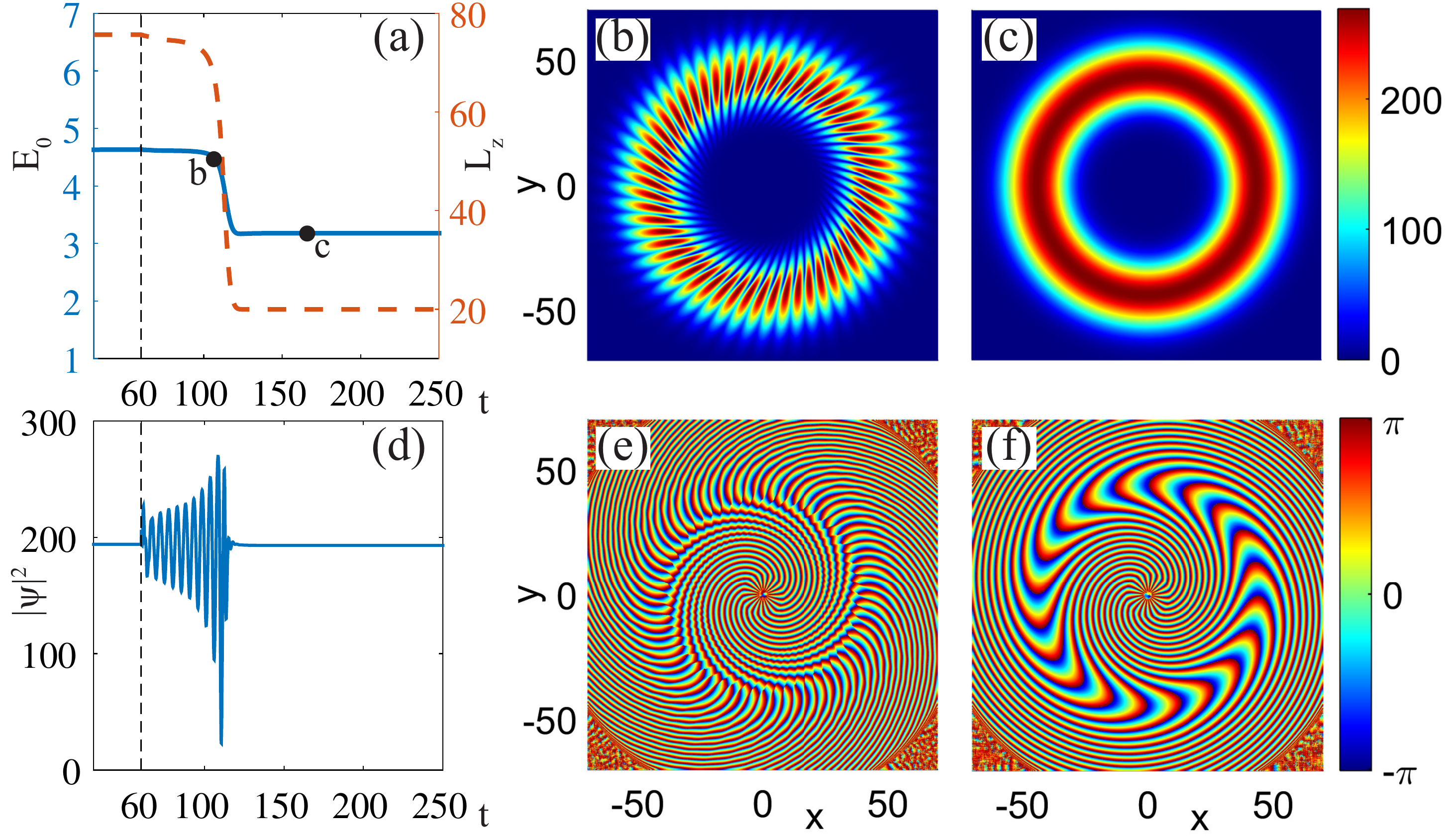}
\caption{(Color online). Evolution of a dynamically (modulationally) unstable state with $m>m_{ds}$ in the regimee $II$. (a) Energy and (normalised) angular momentum evolution during steady state switching triggered by the oscillatory instability; (b)-(c) density and (e)-(f) phase distribution at the (b,e) intermediate and (c,f) final stages of evolution. (d) Peak density evolution at an arbitrary point. Dashed line in (a,d) indicates introduction of a weak pulsed perturbation.}
\label{E_decay}
\end{figure}

Numerical simulations of the full 2D model (\ref{GPE}) with a weak, incoherent perturbation applied to the steady current, show remarkable agreement with the predictions of the 1D stability theory. Indeed, in the regime $II$, for $m>m_{ds}$, the initial stage of the instability development manifests in {\em oscillating and rapidly growing} density perturbations [Fig. \ref{E_decay}(d)], whereby the condensate fragments [Fig. \ref{E_decay}(b)]. Fluctuations around the steady state grow without the formation of  surface modes \cite{Dubessy12,Woo12}, confirming the validity of our $1D$ approximation. During the long-term, nonlinear stage of instability development, the azimuthal flow "heals" [Fig. \ref{E_decay}(c)], and the system attains a new, dynamically stable steady state [Fig. \ref{E_decay}(a)]. Fig.~\ref{E_decay} shows a typical scenario of the oscillatory instability development causing the system to enter a steady state with reduced angular momentum and energy. 

\begin{figure}[h!]
\includegraphics[width=\columnwidth]{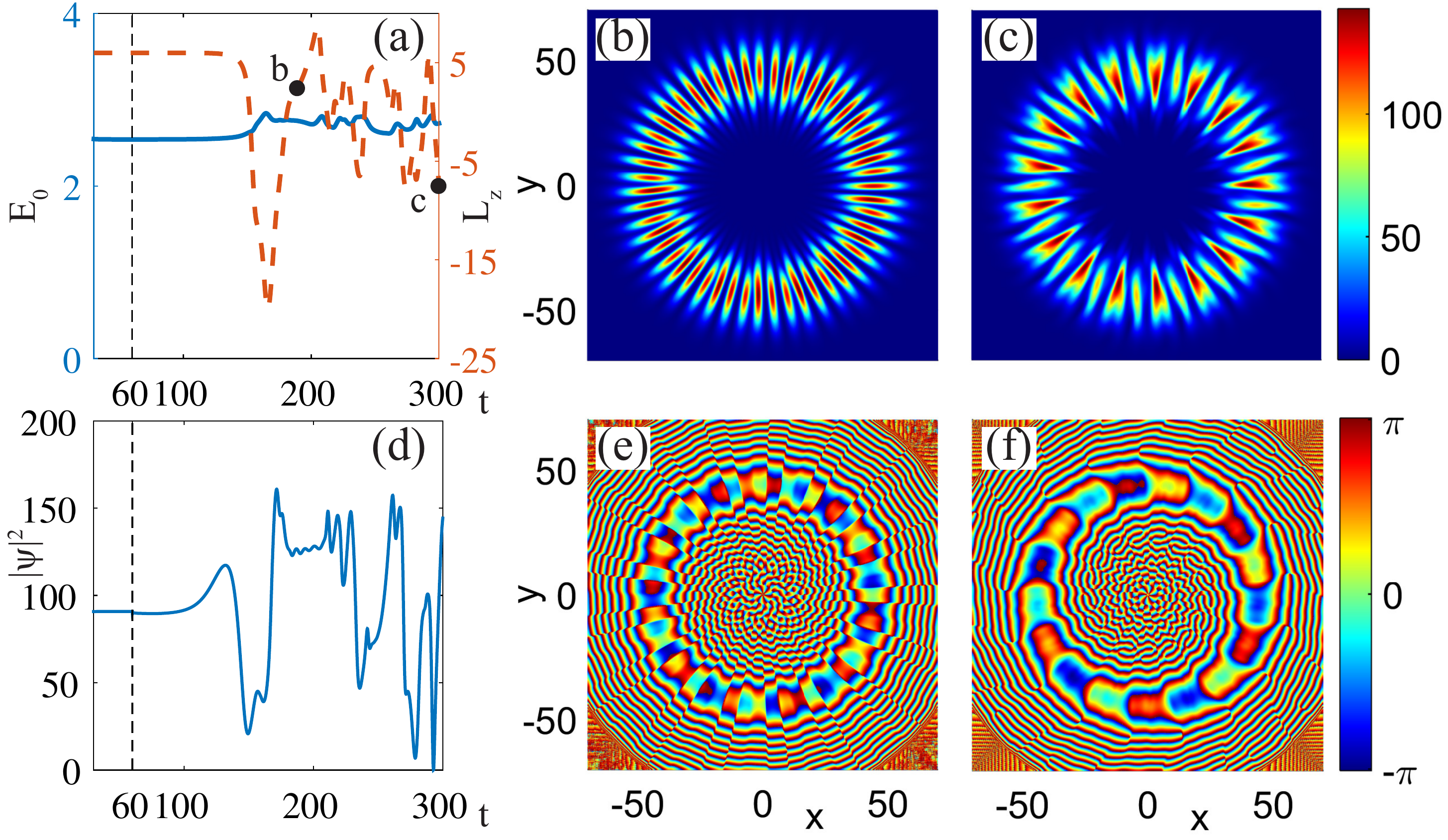}
\caption{(Color online). Evolution of a dynamically (modulationally) unstable state with $m>0$ in the regime $I$. (a) Energy (\ref{Energy}) and (normalised) angular momentum (\ref{Momentum}) evolution during decay of the $m=7$ current triggered by oscillatory instability; (b)-(c) density and (e)-(f) phase distribution at the (b,e) intermediate and (c,f) final stages of evolution. (d) Peak density evolution at an arbitrary point. Dashed line in (a,d) indicates introduction of a weak pulsed perturbation.}
\label{d_decay}
\end{figure}

In contrast, in the regime $I$, where $m_{ds}=0$, once the instability of the persistent current is triggered, the steady flow never recovers [Fig. \ref{d_decay}]. The rate of instability-triggered decay depends on the maximum instability growth rate, ${\rm max}(\Gamma)$, which accounts for the broad transition region from dynamically unstable to stable regime depicted in Fig. \ref{k_vs_m}(d).

{\em Energetic stability.} -- According to the Landau criterion, a superfluid flow without dissipation is no longer energetically favourable above a critical velocity. The flow of a conservative superfluid in a quasi-1D annular geometry becomes {\em energetically unstable} only when its angular velocity exceeds a local speed of sound \cite{Rokhsar97,Javanainen98,Moulder12}. In our variables, this condition can be written in terms of a critical amount of angular momentum carried by the persistent flow as: $m>m_{es}\equiv r_0v_s$, where $v_s=\sqrt{g_cn^0_c}$. 

In contrast to traditional superfluids \cite{Wu03,Modugno04,Baharian13}, the open-dissipative superflow is energetically unstable even in the subsonic regime, since the real part of the excitation spectrum contains negative components for any $m$ \cite{Wouters10,Malpuech14}, thus leading to negative contributions to the energy. As discussed in \cite{Wouters10}, this formal violation of the Landau criterion should mean energetic instability of the dissipative superflow for any velocity. However, in the annular geometry, the typical energetic instability scenario, whereby the persistent current undergoes a series of phase slips reducing its angular momentum (see Fig. \ref{E_decay}), can be observed only in the MI domain for $m>m_{ds}$. Indeed, outside the MI domain $\Gamma(k)<0$, and exponential decay of excitations suppresses the development of instability. The transition rate between any two states with $m$ and $m'=m-k$ due to the time-dependent excitation (\ref{delta_psi}), is defined by the second-order perturbation of the full energy functional, $E=E_0+(i/2)\int[Rn_R-\gamma_c]|\psi|^2rdrd\theta$, and decays as $\sim 1/t$, leading to long lifetimes of the dynamically stable persistent flows. Indeed, for $m \ll m_{ds}$ [red dots in Fig. \ref{k_vs_m}(d)], we do not observe decay of the persistent flow in the 2D numerical simulations even for moderate perturbation amplitudes.

{\em Conclusions. --}  We have analysed the dynamical (modulational) instability of the persistent currents in dissipative polariton condensates confined by all-optical annular traps. Above critical values of orbital angular momentum, the flows suffer from the oscillatory instability, which leads to either dynamical switching to new metastable steady states or destruction of the superfluid flow. The formal non-compliance with the traditional Landau criterion of superfluidity should not impede observation of persistent currents due to suppressed development of energetic instability in dissipative superfluids. The possibility to create a polariton condensate in an optically induced annular trap has already been explored experimentally, and spontaneous formation of vortices and patterns has been observed in such traps \cite{Toroidal_pBEC_11,Toroidal_pBEC,Snoke14}. Provided that coherent imprinting of orbital angular momentum \cite{Sanvitto10} can be realised for these systems, the test of our predictions could be feasible.

{\em Acknowledgements. --} This work was supported by the Australian Research Council (ARC) through the Discovery and Future Fellowship schemes. G.L. acknowledges support of the China Scholarship Council (CSC).


\begin{thebibliography}{99}
\bibitem{units} Throughout the paper we use physical values that are reasonably close to those measured for AlGaAs microcavity polaritons: $g_c=6\times 10^{-3}$ $meV$$\mu m^2$, $g_R=2g_c$, $\gamma_c=0.33$ $ps^{-1}$, and $R=0.01$ $\mu m^2$ $ps^{-1}$. The polariton effective mass entering our model is related to the mass of electron as: $m=10^{-4}m_e$.
\end{thebibliography}
\end{document}